\documentclass[superscriptaddress,aps,pra,twocolumn,floatfix]{revtex4-2}

\usepackage{graphicx} % Required for inserting images
\usepackage{amsmath}
\usepackage{float}
\usepackage{hyperref}

\usepackage{siunitx}
\DeclareSIUnit{\torr}{torr}

\usepackage{gensymb}
\usepackage{xcolor}
\usepackage[normalem]{ulem}

\newcommand{\rb}{\ensuremath{^{87}\text{Rb }}}
\newcommand{\nit}{\ensuremath{\text{N}_{2}\text{ }}} 
\definecolor{mygreen}{rgb}{0,0.5,0}
\definecolor{mygreen2}{rgb}{0,0.85,0}
\definecolor{mygrey}{rgb}{0.5,0.5,0.5}
\definecolor{myred}{rgb}{0.75,0,0}
\definecolor{myblue}{rgb}{0,0,0.75}
\definecolor{mymagenta}{cmyk}{0,1,0,0.12}
\definecolor{mycyan}{cmyk}{1,0,0,0.12}
\definecolor{myorange}{rgb}{1,0.5,0}
\definecolor{myviolet}{rgb}{0.5,0.0,0.75}
\definecolor{mybrown}{cmyk}{0,0.50,1,0.41}

\begin{document}

 \title[Article Title]{Fast, powerful,  low-noise optical pumping of an atomic vapor with semiconductor optical amplifiers}

\newcommand{\ICFO}{ICFO - Institut de Ci\`encies Fot\`oniques, The Barcelona Institute of Science and Technology, 08860 Castelldefels (Barcelona), Spain}
\newcommand{\ICREA}{ICREA - Instituci\'{o} Catalana de Recerca i Estudis Avan{\c{c}}ats, 08010 Barcelona, Spain}
\newcommand{\USP}{Instituto de F\'{i}sica, Universidade de S\~{a}o Paulo, 05315-970 S\~{a}o Paulo, SP-Brazil}

 \author{Diana Méndez-Avalos}
\email{diana.mendez@icfo.eu}
 \affiliation{\ICFO}

\author{Théo Louzada Meireles}
\affiliation{\USP}
\affiliation{\ICFO}

\author{Morgan W. Mitchell}
\affiliation{\ICFO}
\affiliation{\ICREA}

\author{Aleksandra Sierant}
\affiliation{\ICFO}

\begin{abstract}
We use a \rb atomic vapor, suitable for an optically-pumped magnetometer (OPM) in Earth-field conditions, to study the noise properties of three strategies for generating pulsed optical pumping. We compare a frequency-modulated (FM) laser, amplitude modulation (AM) via an acousto-optic modulator (AOM), and amplitude modulation via a semiconductor optical amplifier (SOA). Pumping the ensemble to operate as a Bell-Bloom OPM, and with an equal degree of spin polarization, the three methods give nearly identical sensitivity, showing that the SOA, despite being an active device, can introduce negligible additional noise. Pumping the ensemble to operate as a free-induction-decay OPM, we observe longer unpumped coherence times with the SOA-AM method than with the FM method. Finally, using the higher power available from the SOA, we demonstrate an environment-limited sensitivity of \SI{80}{\femto\tesla\per\sqrt{\hertz}} at \SI{600}{\hertz} and \SI{200}{\femto\tesla\per\sqrt{\hertz}} at \SI{4}{\kilo\hertz}, one to two orders of magnitude beyond what was achievable with the other pumping methods. 

\end{abstract}

\maketitle

% \btext{new suggested text for the manuscript}

% \ctext{sometimes also new text suggested for the manuscript}

% \brtext{old text, to be replaced with blue or removed}

% \mtext{mwm comment}

% \rtext{dma comment}

% \vtext{as comment}

% \gtext{moved paragraph}

% \otext{paragraph in peril}

\section{Introduction}

Optical pumping is a central tool in a wide range of atomic and quantum technologies. Efficient preparation and control of spin polarization is essential in applications such as atomic clocks \cite{Knappe2004,KnappeOL2005,BenedictoPRL2022,HuangThesis2019}, quantum memories \cite{reim2010towards,zhao2009long,sprague2013efficient,lvovsky2009optical}, medical imaging \cite{chupp2000medical,nikiel2007metastability,collier2013high}, and optically pumped magnetometers (OPMs) \cite{Kominis2003,Allred2002,Hunter2018,Jimenez-Martinez2012,Deans2018a,Bevilacqua2016,KorthJGRSP2016,zhang:22}. In many of these systems, it is important to switch between pumping, to generate spin polarization, and not pumping, to allow the spin polarization to evolve with maximum coherence \cite{Jimnez-Martnez2010, Liu2025, Makarov2025}.

OPMs \cite{Fabricant2023} operating in the magnetic field of the Earth can have particularly stringent requirements on switched optical pumping. In this regime, OPMs have the potential to be simultaneously sensitive \cite{LuciveroPRAppl2021, LuciveroPRAppl2022} and accurate  \cite{BennettSensors2021}, making them promising for Earth observation \cite{FRIISCHRISTENSENASR2008, deconinckSPIE2025nanomagsat}, magnetic navigation \cite{MuradogluARX2025}, and geotechnical \cite{Schnau2025} applications. Relevant magnetometer strategies such as Bell-Bloom (BB) \cite{Troullinou2021, SierantARX2026} and free-induction-decay (FID) \cite{Hunter2023} employ stroboscopic optical pumping, i.e., pumping with low duty-cycle pulses spaced by one Larmor period.  At the $\sim$ \SI{500}{\kilo\hertz} Larmor frequencies typical of Earth-field operation, this requires sub-\SI{}{\micro\second} pulse durations. With sufficient pumping, it is also possible to reach the so-called light-narrowing regime \cite{AppeltPRA1999, Zhang2023, Guo2019}, which boosts coherence time and sensitivity. At the same time, optical pumping is a strong perturbation to the spin dynamics, and can introduce both noise and systematic errors into the magnetometer reading, through fluctuations in pulse characteristics including optical frequency, pulse arrival time, and pulse energy. All of these must be well controlled to reach the accuracy and sensitivity potential of the OPM.

Several modulation strategies have been employed for pulsed optical pumping, including frequency modulation (FM) \cite{Bell1961,Troullinou2021,Lipka2024}, amplitude modulation (AM) \cite{wang2016study,Hunter2023,Jimnez-Martnez2012}, and polarization modulation (PM) \cite{fescenko2013bell}. Among these, AM provides a conceptually simple approach in which the laser frequency and polarization are held fixed while the optical power is switched synchronously with the atomic precession.

Here we implement fast, high-power AM optical pumping using a continuous-wave semiconductor laser followed by a gain-modulated semiconductor optical amplifier (SOA) \cite{Buser2024}, a configuration also referred to as master oscillator power amplifier (MOPA). In our case, the laser is a distributed Bragg reflector (DBR) laser and SOA is one designed for high output power, known as a booster optical amplifier (BOA). DBR lasers have been widely employed for optical pumping of OPMs. The laser-SOA approach for generating strong AM pulses at atomic resonance wavelengths is described in \cite{Buser2024}. To our knowledge, ours is the first test of the noise properties of this approach, which is critical for SOA use in precision atomic sensors. 

We demonstrate high-extinction-ratio optical pulses at repetition frequencies up to \SI{500}{\kilo\hertz}, suitable for geomagnetic-field OPMs. To understand the noise contribution of the SOA, we operate a Bell-Bloom OPM using three different pumping modulation strategies: FM, AM using the laser followed by an acousto-optics modulator (AOM-AM), and AM by gain modulation of a SOA following the laser (SOA-AM). As described in \autoref{sec:BOA-OPM} below, with a comparable degree of optical pumping, the three modulation strategies produce nearly identical OPM sensitivity, showing the SOA is not introducing extra noise. Moreover, the SOA when operated at higher powers (not available to the other approaches), achieves sensitivity of \SI{80}{\femto\tesla\per\sqrt\hertz}, about an order of magnitude better than the others.  Finally, we observe FID relaxation of the spin ensemble when pumped with the SOA-AM and FM approaches, and directly observe a longer coherence time with SOA-AM pumping, which we attribute to fluctuating light shifts and off-resonance scattering in the FM approach, which is absent in the SOA-AM approach, due to the strong attenuation by the SOA in its off state.

\section{Spin system, pumping and probing}
\label{sec:SpinPumpProbeMethods}

To test the degree to which the  modulation schemes introduce noise and relaxation into a coherently-precessing atomic ensemble, we implement Bell-Bloom (BB) excitation of the spin polarization \cite{Bell1961}. That is, we send one low-duty-cycle pump pulse per Larmor precession cycle. We also implement Faraday-rotation monitoring of the spin precession using an off-resonance probe beam.  

\begin{figure}[h!]
    \centering

\includegraphics[width=\columnwidth]{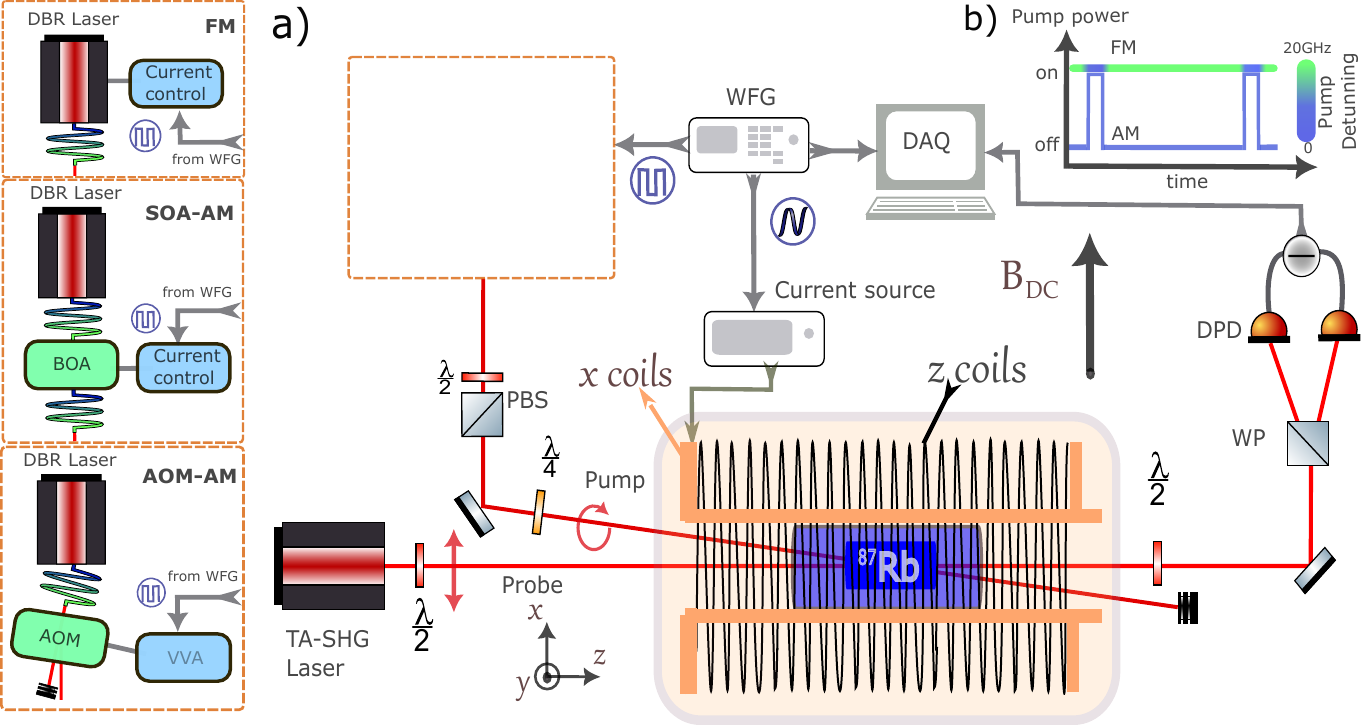}
\caption{(a) Bell–Bloom OPM setup. \rb vapor cell, heater, and coils are enclosed within a three-layer mu-metal shield. The three different pump laser sources are presented in the orange dotted boxes: the frequency modulation (FM) scheme is implemented by controlling the current injected into the laser; the semiconductor optical amplifier-based amplitude modulation (SOA-AM) scheme is implemented by modulating the current injected into the BOA; while for the acousto-optic modulator-based amplitude modulation (AOM-AM) scheme, the modulation is implemented by controlling the voltage of the VVA that controls the attenuation of the 1st-order diffracted beam. BOA - booster optical amplifier, AOM - acousto-optic modulator, VVA - variable voltage attenuator, WP - Wollaston prism, DPD - differential photodetector, PBS - polarized beam splitter, DAQ - data acquisition card, TA-SHG - tapered amplified second harmonic generator, WFG - waveform generator, DBR - distributed Bragg reflector. (b) Bell–Bloom optical pumping modulation schemes. Frequency modulation (FM): once per cycle, the pump laser frequency is swept from 20 GHz blue-detuned to resonance with the $F=1$ hyperfine transition and back. Amplitude modulation (AM): the pump laser remains on resonance while its optical power is modulated from zero to maximum once per cycle. }
    \label{fig:schemeandfullsetup}
\end{figure}

The experimental system is shown schematically in Fig.~\ref{fig:schemeandfullsetup}: A \SI{3}{\centi\meter}-long glass  cell containing isotopically enriched \rb and \SI{100}{\torr} of \nit buffer gas is heated to \SI{105}{\celsius} to achieve an atomic density of \SI{8.2e12}{atoms\per\centi\meter\cubed}. A constant magnetic field along the $x$-axis is generated using a low-noise current source (Twinleaf CSUA300). 
A linearly-polarized probe beam, propagating along \textit{z} axis, detuned by \SI{20}{\giga\hertz} to the blue of the \rb D$_1$ transition experiences Faraday rotation and is detected with a balanced polarimeter.  The circularly-polarized pump beam  propagates through the cell at a small angle in the same direction as the probe. For additional details, see \cite{Troullinou2021}.

As the probe passes through the atoms, it experiences Faraday rotation, in which the plane of linear polarization rotates by an angle proportional to the spin polarization \cite{Budker2013}. The Faraday rotation signal voltage is given by \cite{Kong2020}
\begin{equation}
\label{eqn:SignalAmplPol}
    V(t) = GRP_{\text{pr}}\sin( 2\theta_F(t)) + N(t),
\end{equation}
where $V(t)$ is the differential photodetector (DPD) voltage as a function of time $t$, $G$ and $R$ are, respectively, the detector gain and responsivity, $P_{\text{pr}}$ is the total probe power incident on the detector, $N(t)$ is the polarization noise, $\theta_F(t)$ is the Faraday rotation angle and $\theta_F (t) \propto \mathcal{P}_z(t)$, with $\mathcal{P}_z$ the mean atomic spin polarization projected onto the probe axis. 
%$\theta_F(t)=\pi r_ecf_{D_1}nl/\Delta_{\text{probe}}*\mathcal{P}_x(t)$
\cite{Seltzer2008}. 

We infer the instantaneous quantity \(\mathcal{P}_z(t)\) using 
\cite{Seltzer2008}
\begin{equation}
    \theta_F(t)=\frac{l n r_e c f_{D_1}}{2\Delta} \mathcal{P}_z(t),
\end{equation}
where $l$ is the vapor cell length, $n$ is the atomic number density, $r_e$ is the classical electron radius, $c$ is speed of light, $f_{D_1}$ is the oscillator strength of the $D_1$ \rb line and $\Delta$ is the probe detuning. For BB-OPM or FID-OPM operation, we fit \(\mathcal{P}_z(t)\) with a sinusoid or a sinusoid with a decaying exponential envelope, respectively, to obtain the fitted maximum oscillation amplitude \(\mathcal{P}_0 = \max_{t} |\mathcal{P}(t)| \), which we will refer to as the degree of polarization (DOP). Equal DOP implies equal efficiency in optical pumping, indpendently of the optical pumping method. % \otext{Two measurements performed under identical probe and detection conditions and giving the same fitted Faraday-rotation amplitude correspond to the same spin polarization degree.} \mtext{mwm: I don't understand this last sentence. I would think ``giving the same fitted Faraday-rotation amplitude'' would be enough. Why does ``probe and detection conditions'' appear here?} \vtext{because if you would have different probe powers or gain on the detector you would have different amplitude - or you think this is obvious and we dont need to specify?} \mtext{mwm: if you agree with what I wrote about calculating $\mathcal{P}_z$, then I think the recipe is there and it is not necessary to say ``for equal probe ($P_\mathrm{in}$) and detection ($G$, $R$).'' In fact the recipe would work also for unequal values of these parameters.}

%\begin{equation}
 %   \theta_F(t)=\frac{l n r_e c f_{D_1}}{2\Delta} \mathcal{P}_z(t),
%\end{equation}
%where $l$ is the vapor cell length, $n$ is the atomic number density, $r_e$ is the classical electron radius, $c$ is speed of light, $f_{D_1}$ is the oscillator strength of the \rb D\textsubscript{1}  line, $\Delta$ is the probe detuning,  \mtext{mwm: do we use these equations for something? For what? }

% \section{Pump modulation methods}
% \label{sec:characterization}
\begin{figure}[t]
    \centering  
        \includegraphics[width= \columnwidth]{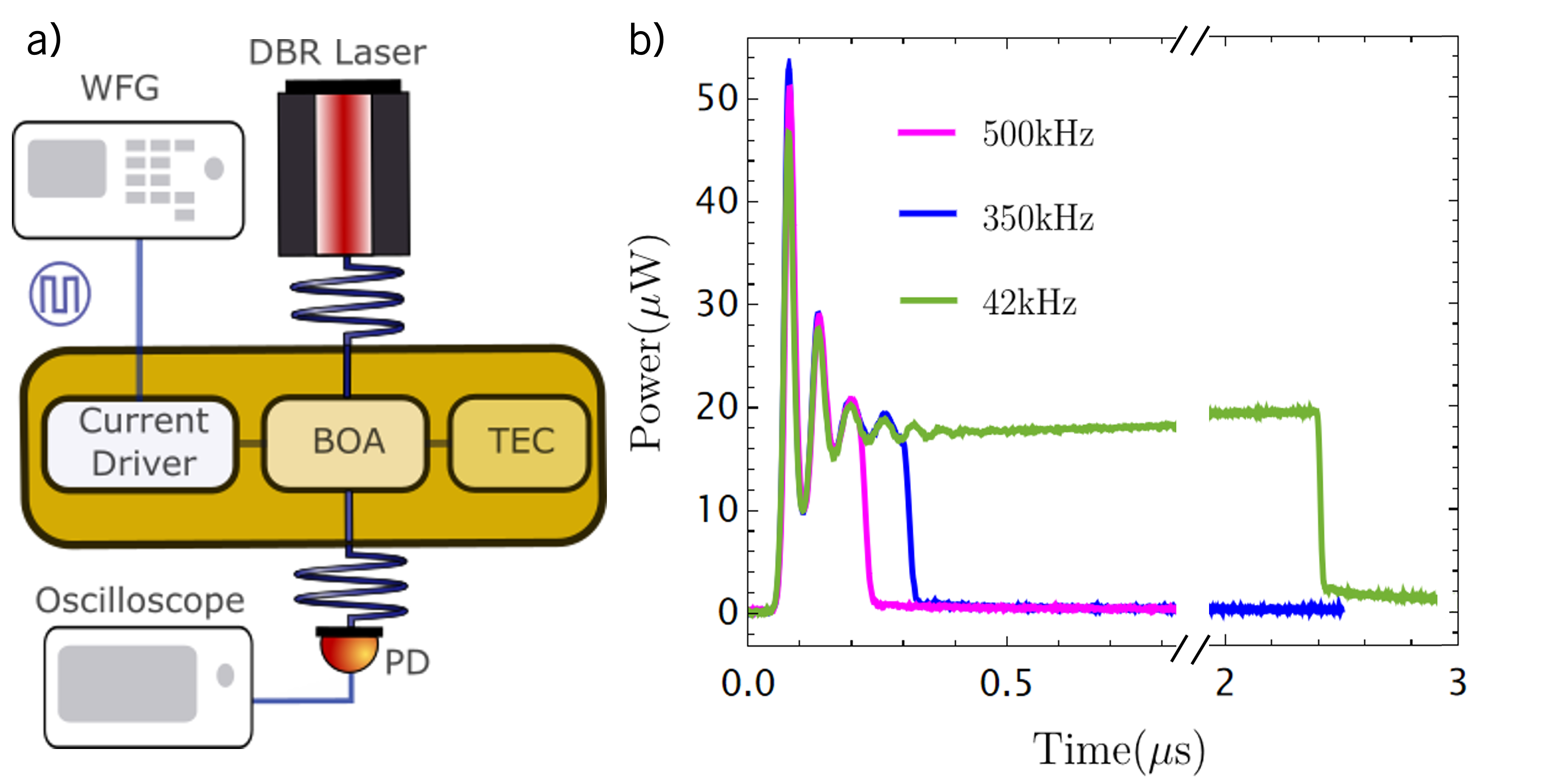}
    \caption{a) BOA characterization setup. A waveform generator (WFG) provides the modulation signal to the BOA current driver. The BOA is seeded by a distributed Bragg reflector (DBR) laser, and its temperature is stabilized using a thermoelectric controller (TEC). The amplified and modulated light is sent directly to a photodetector (PD) for pulse characterization. b) Optical pulses generated by the BOA and detected at PD when BOA is driven by nominally-rectangular unidirectional current pulses (low of \SI{0}{\milli\ampere}, high of \SI{199.4}{\milli\ampere})  at 10$\%$ duty cycle at repetition frequencies of \SI{500}{\kilo\hertz} (magenta), \SI{350}{\kilo\hertz} (blue), and \SI{42}{\kilo\hertz} (green). The latter  SOA-AM driving condition was used for the low-power magnetometry comparison of Fig.~\ref{fig:sensitivities}. Each trace represents an average over 64 pulses.}
    \label{fig:setupandpulses}
\end{figure}

%\vtext{suggestion:}\btext{As the probe passes through the atoms, it experiences Faraday rotation proportional to the spin polarization projected onto the probe axis. We denote this time-dependent projection by \(\mathcal{P}_z(t)\). The differential photodetector voltage is
%\begin{equation}
%\label{eqn:SignalAmplPol}
%    V(t) = GRP_{\mathrm{in}}\sin 2\theta_F(t) + N(t),
%\end{equation}
%where \(G\) and \(R\) are the detector gain and responsivity, respectively, \(P_{\mathrm{in}}\) is the total probe power incident on the detector, \(N(t)\) is the detection noise, and \(\theta_F(t)\) is the Faraday rotation angle. In the following, the reported spin polarization degree is not the instantaneous quantity \(\mathcal{P}_z(t)\), but the fitted oscillation amplitude \(\mathcal{P}_0\). It is obtained by fitting the demodulated or time-domain Faraday signal to the corresponding spin-precession model, with the calibration between \(\theta_F\) and \(\mathcal{P}_z\) calculated from the atomic density, cell length, probe detuning, and optical parameters. \vtext{Dianita, is this accurate?} Thus, two measurements performed under identical probe and detection conditions and giving the same fitted Faraday-rotation amplitude correspond to the same spin polarization degree.}

The three methods of optical pumping are implemented as follows: 
The FM method replicates the conditions in \cite{Troullinou2021}. Briefly, the DBR (Photodigm DBR, 794.978DBRH-MHF-TO8) laser frequency is modulated in a square-wave pattern so that it is on-resonance with the $F=1$ hyperfine state of  the \rb D$_1$ transition for $10\%$ of the duty cycle and \SI{20}{\giga\hertz}  blue-detuned off-resonance for the rest. The temperature and injection current of the diode are controlled by ThorLabs LDC202C and TED200C, respectively. The frequency modulation is accomplished by modulating the DBR current, controlled via a waveform generator (WFG, Keysight, 33500B). Due to this current modulation, the laser power is \SI{17.4}{\percent} lower when off resonance. In the FM scheme [Fig.~\ref{fig:schemeandfullsetup}(a)], light from the DBR laser is sent directly to the atomic vapor. The SOA is not used in this method. 

In the AM methods, the pump laser is held continuously on resonance with the \rb D$_1$ transition, addressing  the $F=1$ hyperfine state while its optical power is modulated by controlling the attenuation of an AOM or by controlling the BOA gain. %The laser output is fiber-coupled into the BOA, which is driven to alternately amplify and attenuate the light, thereby producing the pumping pulses.  

In the AOM-AM scheme, the AOM (IntraAction ATM-3501A2) is driven by a voltage-controlled oscillator (VCO, Mini-Circuits ZOS-535+) at \SI{80}{\mega\hertz} (chosen to maximize first-order diffraction efficiency), the output of which is attenuated by a voltage-variable attenuator (VVA, Mini-Circuits ZX73-2500M-S+) before being amplified by a fixed-gain power amplifier. The WFG controls the VVA attenuation, and thus the optical power in the first-order diffracted beam, which is used for optical pumping. %\vtext{suggestion:} \btext{The VVA provides approximately \SI{40}{dB} RF attenuation, but the optical extinction ratio of the AOM-AM system depends additionally on RF leakage, scattering, beam separation, and spatial filtering.}  
For typical beam sizes, such systems provide extinction ratios on the order of \SI{40}{dB}, while optimized AOM-based systems have demonstrated values near \SI{60}{dB}~\cite{macrae2021optical,wojtewicz2018response}. %\mtext{mwm: am I correct this is an estimate and not a measured value?  If an estimate, where is it from?  } } \vtext{its an estimate i think, based on the extinction ratio for the VVA, not an optical extinction ratio. there are few papers that report measured optical extinction ratio for AOMs, with better VVAs they report 58-60 dB of attenuation, but not sure if it is relevant to cite them (different systems, different wavelength and better VVA). There are also companies that sometimes report optical extinction ratio for their AOMs, usually 40-65 dB (the AOM used in this measurement does not specify that, unfortunetly)} %The joint VVA/AOM combination modulation bandwidth is at the order of tens of kHz \rtext{??} \vtext{i would say tens of kHz based on the estimation from the specified rise and fall times for VVA - they give 6 and 25 kHz, but in principle you could do faster just not high power... so hard to say}}. 
No SOA is used in this method.  

In the SOA-AM method, the laser is ``chopped'' into pulses with the SOA.  Fig.~\ref{fig:setupandpulses}a) shows the experimental setup used for SOA-AM, which employs SOA  (Thorlabs BOA795P) which is integrated with single-mode fiber in- and out-coupling optics in a butterfly package. The injection current is supplied by a current driver (Koheron DRV200-A-400) and temperature stabilization is provided by a TEC controller (Meerstetter Engineering TEC-1091). The current driver has a nominal \SI{3}{\decibel}  modulation bandwidth of \SI{6}{\mega\hertz}, and is controlled by a WFG with \SI{10}{\percent} duty cycle pulses. A fraction of the pump light is directed to a photodetector (PDA10A-EC), calibrated against a power meter (Thorlabs, PM100D) and recorded using an oscilloscope. 

Fig.~\ref{fig:setupandpulses}b) shows the pulses generated by the BOA when seeded with  \SI{13.8}{\micro\watt} of laser power and driven by the WFG at pulse repetition frequencies of \SI{42}{\kilo\hertz}, \SI{350}{\kilo\hertz} and \SI{500}{\kilo\hertz}, corresponding to \rb Larmor frequencies for fields of \SI{6}{\micro\tesla}, \SI{50}{\micro\tesla} and \SI{71.4}{\micro\tesla}, respectively, and thus more than covering the \SIrange{30}{70}{\micro\tesla} Earth field range. The observed ringing behavior appears to be due to the limits in the current driver bandwidth of \SI{6}{\mega\hertz}.  However, we note that using faster current drive electronics, SOAs on the Rb D lines have demonstrated ON/OFF switching times as short as \SI{50}{\nano\second} \cite{Buser2024}. 

%\brtext{The observed ringing behavior is \mtext{appears to be? how well do we know where this ringing comes from?  how do we know where it comes from?}\rtext{dma: we see it directly when driving the current driver with the wfg and no boa} due to the use of a low-noise current driver (with correspondingly low modulation bandwidth) and parasitic effects in the BOA mount. We have thus not reached the intrinsic speed limits of BOA modulation, which would arise due to carrier dynamics on sub-\SI{}{\nano\second} time scales \mtext{what I see in Buser is that the photon time-scale is picoseconds while the carrier time-scale is ns. It is a very short discussion; I don't understand it, and it doesn't seem crucial to anything we are doing here. I would drop the ps mention. (I'm also quite sure that carrier dynamics is faster than ns. If it were not, how could you have a multi-GHz photodetector?)}. Using faster current drive electronics, SOAs on the Rb D lines have demonstrated ON/OFF switching times as short as \SI{50}{\nano\second} \cite{Buser2024}. } 

\newcommand{\supin}{^{(\mathrm{in})}}
\newcommand{\supout}{^{(\mathrm{out})}}

% \subsection{Coherence ``in the dark''}
% \label{sec:Dark}

\begin{figure*}[t]
    \centering
    \includegraphics[width=\linewidth]{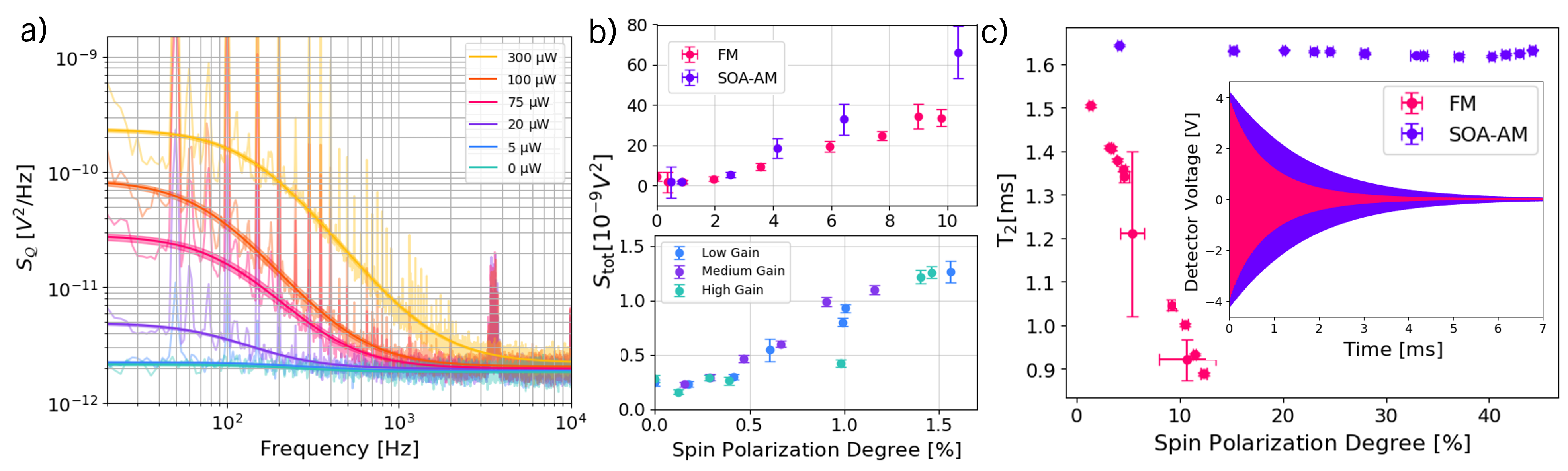}
    \caption{
Comparison of Faraday-rotation noise properties with FM and SOA-AM pumping modulation strategies. (a) SOA-AM's demodulated polarization-noise PSD for different average pulsed pump powers at \SI{70}{\micro\watt} of probe power measured before the atoms and pump duty cycle of 10$\%$. Faded curves show raw data; solid curves show fits to equation.\ref{eqn:NoiseSpectra} obtained using a maximum-likelihood estimator, following \cite{Lucivero2017}, excluding line noise at \SI{50}{\hertz} and its harmonics, and the \SIrange{3}{4}{\kilo\hertz} band. The shaded region indicates the uncertainty estimated via bootstrapping (500 iterations). 
(b) Total spin noise spectral density, fitted with Eq.~\ref{eqn:NoiseSpectra}, as a function of spin polarization degree, fitted to the raw data with Eq.~\ref{eqn:SignalAmplPol} at \SI{60}{\micro\watt} of probe power. %\mtext{haven't said how this is measured, nor how it is defined.}
%\btext{The total noise was fitted with equation \ref{eqn:NoiseSpectra}, while the spin polarization degree was done by fitting the raw data with equation \ref{eqn:SignalAmplPol}} . 
Upper, comparison for FM (in pink) and SOA-AM (in purple). Lower, total spin noise versus DOP for injection currents of \SI{100}{\milli\ampere} (blue, ``low gain''), \SI{160}{\milli\ampere} (violet, ``medium gain'') and \SI{242.7}{\milli\ampere} (green, ``high gain''), with an input seed of $\sim$\SI{1}{\milli\watt}.  c) Fitted coherence time ($T_2$) for free-induction-decay (FID) signals vs. the spin polarization degree for FM (in pink) and BOA (purple) with \SI{70}{\micro\watt} of probe power. Inset: comparison of BOA-based AM and FM FIDs for \SI{15.1}{\percent} and \SI{12.3}{\percent} spin polarization respectively at \SI{70}{\micro\watt} of probe power.}  
    \label{fig:totalPolarization}
\end{figure*}

\section{modulation noise in high-sensitivity magnetometry} 
\label{sec:BOA-OPM}

To make a controlled comparison of the noise performance of the three modulation schemes, we use Eq.~\ref{eqn:SignalAmplPol} to infer the spin polarization from the signal amplitude. Since the detection system and experimental conditions described in \autoref{sec:SpinPumpProbeMethods} are identical across all schemes, two signals of equal amplitude correspond to equal degrees of spin polarization, regardless of the pumping scheme used to achieve them. The pump power of each modulation scheme is then adjusted to produce equal Faraday rotation signal amplitudes, ensuring a fair comparison of noise performance at equivalent polarization levels. 

%\brtext{\btext{From equation \ref{eqn:SignalAmplPol} we can infer that if the detection system and experimental conditions  described in \autoref{sec:SpinPumpProbeMethods} are the same, two signals of the same amplitude will have the same DOP regardless the pump scheme it has been used.  }To compare the noise performance of our three modulation strategies, we use \sout{always} the probe \btext{and experimental} conditions, apply one of the modulation strategies to generate  Larmor-resonant pulses, i.e., BB excitation, and tune the pump power\mtext{ (?) other parameters also? how do you actually tune the system to have equal polarization?} to generate an equal amplitude of Faraday rotation signal \mtext{ what is the resulting amplitude of the FR signal?}.} 

The signal from the DPD, acquired by the data acquisition card (DAQ), is digitally demodulated with the phase reference taken from the  WFG that drives the modulation. Following the methodology of \cite{Troullinou2021}, we compute the sensitivity, i.e., the equivalent magnetic noise spectrum, as 
\begin{equation}
\label{eqn:sensitivity}
    S_B(f)=\bigg(\frac{d\mathcal{Q}}{dB}\bigg)^{-2}\frac{S_{\mathcal{Q}}(f)}{|\hat{R}(f)|^2},
\end{equation}
where ${d\mathcal{Q}}/{dB}$ is the slope of the $\mathcal{Q}$ quadrature signal, $|\hat{R}(f)|^2$ is the normalized frequency response of the spins to a known modulation of the $B$ magnetic field, with   $R(f) = [ d\mathcal{Q}(f)/dB(f) ] / [ d\mathcal{Q}(0)/dB(0) ]$ , and $S_{\mathcal{Q}}(f)$ is the observed noise in the lock-in quadrature component $\mathcal{Q}$, \textit{i.e.}, 
\begin{equation}
\label{eqn:NoiseSpectra}
    S_{\mathcal{Q}}(f)=S_{N} + \frac{\Delta f^2}{f^2+\Delta f^2} S_\sigma,
\end{equation}
where $\Delta f$ is the magnetic-resonance bandwidth, $S_{N}$ the photon shot noise contribution, and $S_\sigma$ the strength of the spin noise contribution. The total atomic noise is $S_{\text{tot}} \equiv \int df\,(S_{\mathcal{Q}}(f)-S_{N}) = S_\sigma \Delta f\pi/2$.

The demodulated polarization-noise power spectral density (PSD) under the SOA-AM scheme, is shown in Fig.~\ref{fig:totalPolarization}a) for different average pulsed pump powers. The spectra are well described by Eq.~\ref{eqn:NoiseSpectra}, with spin noise contribution dominating at low frequencies and photon shot noise (PSN) forming a white-noise floor at high frequencies. From the fits, the dependence of the total atomic noise spectral density on the signal amplitude for different SOA gain settings and comparison with FM pumping is shown in  Fig.~\ref{fig:totalPolarization}b). The DOP reported on Fig.~\ref{fig:totalPolarization} was fitted with Eq.~\ref{eqn:SignalAmplPol}, with the parameters of the experimental conditions from \autoref{sec:BOA-OPM}  As shown in the upper panel of Fig.~\ref{fig:totalPolarization}b, at low DOP, SOA-based AM does not introduce additional noise relative to FM, nevertheless at higher DOP the total noise of the SOA-AM increases around \SI{50}{\percent} relative to FM. As shown in Fig.~\ref{fig:totalPolarization}c), with the increase of the DOP, the coherence time for SOA-AM remains the same within the statistical error. This lacking of decoherence yields to higher sensitivity for the SOA-AM, which would make it more susceptible to the environmental noise. %\mtext{I'm not understanding this.  When I look at Fig. 3b, it looks to me like FM has lower noise (lower $S_{tot}$). Is that what you meant ? } 
In addition, the total noise is  unchanged between different SOA gain settings within the error and increases slightly with stronger pumping, as shown in the lower panel of Fig.~\ref{fig:totalPolarization}b).

%\mtext{I don't see an explanation of how the spin polarization degree is measured/calculated.  }

The observed total spin noise $S_\mathrm{tot}$ includes both intrinsic spin-projection noise (SPN), originating in spin uncertainty relations and the fluctuation-dissipation theorem \cite{MouloudakisPRA2022, MouloudakisPRA2024}, and magnetic noise acting on the polarized spins.  The SPN component is expected to decrease somewhat with increasing DOP, as the system grades from a thermal state to a coherent spin state \cite{Lipka2024}. Consequently, we interpret the observed rise of $S_\mathrm{tot}$ with DOP as an increased sensitivity to magnetic noise; as the spin polarization increases, the magnetometer responsivity also increases, so magnetic noise is more strongly converted into the measured signal. 
%\mtext{can we estimate how much $S_\mathrm{tot}$ should increase with DOP?}\rtext{dma: interesting question, cause I think it $S_\mathrm{tot}$  should decrease with the increasing on polarization, but I think $S_\mathrm{tot}$ increases cause the magnetometer is more sensitive to technical noise from the current sources}

\begin{figure}
    \centering
    \includegraphics[width=0.9\linewidth]{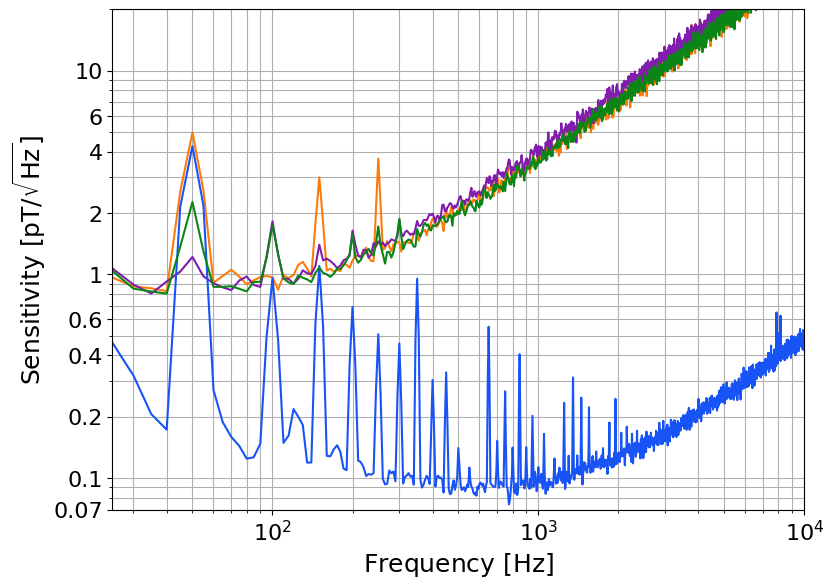}
   \caption{
Magnetic-field sensitivity for BB-OPM, calculated using Eq.~\ref{eqn:sensitivity}.  Orange, purple and green curves show FM, AOM-AM and SOA-AM pumping, respectively, with pumping duty cycle \SI{10}{\percent} and generating equal spin polarization (see text). Blue curve shows sensitivity for SOA-AM pumping with \SI{1}{\percent} duty cycle higher pump power.  
%For AOM- and BOA-based AM (respectively in purple and green) the average pulsed pump power at 10\% duty cycle was set to \SI{2}{\micro\watt} while FM (in orange) was \SI{30}{\micro\watt} . The probe power was \SI{250}{\micro\watt}.  In BOA-based AM at high pump power (in  blue), average power at 1\% duty cycle was \SI{156}{\micro\watt}  (equivalent to \SI{1.44}{\milli\watt} at 10\% duty cycle) \mtext{I don't understand this. In what sense ``equivalent'' ?  Why are these 1 percent and 10 percent numbers not different by a factor of ten ?  } . Probe power was set to \SI{40}{\micro\watt} ,\SI{20}{\giga\hertz} detuned to the blue of the D\textsubscript{1} transition, while the pump was resonant with the $F=1  D_1$ transition}
}
    \label{fig:sensitivities}
\end{figure}

Fig.~\ref{fig:sensitivities} shows the measured magnetic-field sensitivities for FM, AOM-AM and SOA-AM at the highest DOP achievable with the FM technique, and also the SOA-AM at higher DOP due to the higher  pump power available from the SOA.  For the low-DOP conditions, all configurations achieve sub-\SI{} {\pico\tesla\per\sqrt\hertz} sensitivity, consistent with the results reported in \cite{Troullinou2021}. No additional technical noise is observed for the SOA-AM modulation, FM, or AOM-AM implementations.  At higher pump powers, unreachable with the other methods, the SOA-AM equivalent magnetic noise reaches \SI{80}{\femto\tesla\per\sqrt\hertz} at about \SI{600}{\hertz}, about one order of magnitude better than the best FM or AOM-AM sensitivity of \SI{800}{\femto\tesla\per\sqrt\hertz} at about \SI{35}{\hertz}. In the shot-noise-limited regime above about \SI{2}{\kilo\hertz}, the SOA-AM approach is about two orders of magnitude more sensitive.  

At these higher powers, the SOA-AM BB-OPM sensitivity is limited by a low-frequency contribution with a bandwidth comparable to the spin-relaxation rate. The exact origin of this noise we leave for further investigation. We note, however, that it is comparable in strength to the expected magnetic noise in the test chamber: the specified \SI{1}{\nano\ampere\per\sqrt\hertz} current-noise floor of the current source (TwinLeaf, CSUA300) \cite{TwinleafCSBA} times the  (\SI{80}{\nano\tesla\per\milli\ampere}) coil calibration factor gives a magnetic noise level of \SI{80}{\femto\tesla\per\sqrt\hertz}, similar to the observed noise floor. 

\section{residual pumping and spin coherence}

None of the modulation methods used here are guaranteed to reduce the pumping to zero in the nominally ``OFF'' periods. The FM method detunes the pump light, which reduces but does not eliminate pump-induced scattering. The AOM-AM method employs diffraction of gaussian beams, which does not perfectly separate them. The SOA in its OFF state has no detectable laser transmission, but may still have an output due to spontaneous emission, if the injection current remains finite. 

To assess the effect of these imperfections on spin coherence, we performed a free-induction-decay (FID) measurement. As in the BB-OPM tests described in the preceding section, we use stroboscopic optical pumping to generate spin polarization, switch the pump to the corresponding ``OFF'' condition, and use Faraday rotation to observe the spin precession as it relaxes toward zero. Representative signals are shown in Fig.~\ref{fig:totalPolarization}c) (inset). Fitting these with Eq.~\ref{eqn:SignalAmplPol} and the FID function
\begin{equation}
    \mathcal{P}_z(t)=\mathcal{P}_0\sin(\omega_L t+\phi_0)e^{-t/T_2}
\end{equation}
we obtain the initial DOP, $\mathcal{P}_0$, and the transverse spin relaxation time  $T_2$. Adjusting the DOP by tuning the pump laser power, we observe, see Fig.~\ref{fig:totalPolarization}c), a clear trade-off between signal strength and coherence with the FM method, but no such trade-off with the SOA-AM method. We conclude that the SOA-AM method avoids the ``OFF''-state depolarization that is seen in the FM method.

\section{Conclusions}

We have demonstrated that a laser-semiconductor optical amplifier combination can be used as an
effective tool for high-bandwidth amplitude modulation of resonant pump light, enabling precise optical pumping over a broad dynamical range. 
By modulating the SOA gain while seeding it with a narrow-linewidth laser, high-extinction optical pulses can be generated without perturbing the laser frequency or coherence.

The technique was evaluated in a Bell–Bloom optically pumped magnetometer, where SOA-based amplitude modulation was directly compared with  frequency modulation and AOM-based amplitude modulation. Under comparable operating conditions, the SOA introduces negligible additional technical noise. The magnetometer performance remains limited by spin-projection noise at low frequencies and by photon shot noise at high frequencies. At higher pump powers, the SOA reaches a higher degree of spin polarization and a sensitivity of \SI{80}{\femto\tesla\per\sqrt\hertz}, limited by environmental noise conditions.  We show also that the SOA-AM approach preserves coherence during the un-pumped periods better than the FM approach. 

These results show that gain-modulated semiconductor optical amplifiers can generate fast, low-noise, high-power optical pumping pulses with high extinction. Such capabilities are particularly relevant for applications that require precise temporal control of optical pumping over a wide range of intensities, including atomic sensors, atomic clocks, and quantum interfaces based on atomic ensembles.

\section{Acknowledgments}
We thank G. Buser, C. Kiehl and S. Tabares Giraldo for helpful discussions. The work was funded by the European Commission projects Field-SEER (ERC 101097313), OPMMEG (101099379) and QUANTIFY (101135931), the Spanish Ministry of Science MCIN project SAPONARIA (PID2021-123813NB-I00) and SALVIA (PID2024-158479NB-I00), ``Severo Ochoa'' Center of Excellence CEX2019-000910-S;  Generalitat de Catalunya through the CERCA program,  DURSI grant No. 2021 SGR 01453 and QSENSE (GOV/51/2022).  Fundaci\'{o} Privada Cellex; Fundaci\'{o} Mir-Puig. DMA acknowledges funding from the European Union’s Horizon Europe research and innovation programme under the MSCA Grant Agreement No. 101081441.
%\bibliographystyle{biblio/apsrev4-1no-url-title}
%\bibliography{main}
%merlin.mbs apsrev4-1.bst 2010-07-25 4.21a (PWD, AO, DPC) hacked
%Control: key (0)
%Control: author (72) initials jnrlst
%Control: editor formatted (1) identically to author
%Control: production of article title (0) allowed
%Control: page (0) single
%Control: year (1) truncated
%Control: production of eprint (0) enabled
%

\end{document}